\pdfoutput=1 
\documentclass[aps,twocolumn,prl,showpacs,amsmath,amssymb,10pt]{revtex4-1}
\usepackage{graphicx}
\usepackage{scrextend}
\usepackage{manfnt,pifont}

\usepackage{url}

\usepackage{CJK}

\begin{document}

\begin{CJK*}{GB}{}
\CJKfamily{gbsn}


\title{Mellin amplitudes for  $AdS_5\times S^5$}

\author{Leonardo Rastelli}
\author{Xinan Zhou (ÖÜÏ¡éª)}
\affiliation{C. N. Yang Institute for Theoretical Physics\\ Stony Brook University\\ Stony Brook, NY 11794, USA}

\date{\today}

\begin{abstract}

We revisit the calculation of holographic correlation functions in IIB supergravity on  $AdS_5\times S^5$. Results for four-point functions simplify drastically when 
expressed in Mellin space.
We  conjecture a compact formula for the four-point functions of one-half BPS single-trace operators of arbitrary weight.
Our methods  rely on general consistency conditions  and eschew detailed knowledge of the supergravity effective action.

\end{abstract}



\maketitle
\end{CJK*}

\noindent
{\bf Introduction.} Despite almost two decades of relentless efforts, we are still far from harnessing the full computational power of the AdS/CFT correspondence. In 
the canonical duality \cite{Maldacena:1997re, Gubser:1998bc, Witten:1998qj} between ${\cal N}=4$ super-Yang Mills (SYM) theory and IIB string theory on  $AdS_5\times S^5$, the bulk description is 
most tractable in the classical supergravity regime,
which describes  planar SYM theory at large 't Hooft coupling. 
Supergravity is  however still a complicated non-linear theory,
and only the simplest observables have been computed so far. In this letter we revisit the holographic calculation of four-point correlation functions of 
one-half BPS single-trace operators \footnote{Three-point correlators of arbitrary one-half BPS operators were computed in the early days of AdS/CFT \cite{Freedman:1998tz, Lee:1998bxa} and found to agree with the tree-level SYM results, a fact subsequently explained by a non-renormalization theorem, see, {\it e.g.}, \cite{Baggio:2012rr} and references therein.}.
 In the supergravity limit, there is a straightforward algorithm that  computes them as a sum of tree-level Witten diagrams, 
whose vertices are encoded in the $AdS_5$ effective action \cite{Arutyunov:1999fb} obtained by Kaluza-Klein (KK) reduction of IIB supergravity on $S^5$. 

The difficulty of the calculation grows quickly with the KK level  
and complete results are only available for a handful of four-point correlators.
There are some hints that  final answers are simpler than the  intermediate calculations. For example, evaluating the four-point function of
the lowest KK mode (corresponding to the stress-tensor supermultiplet) is a non-trivial task \cite{DHoker:1999pj, Arutyunov:2000py}, but the result
can be written as a single quartic Witten diagram \cite{Dolan:2004iy}. 
One is tempted to draw an analogy
with tree-level gluon scattering amplitudes in $4d$ Yang-Mills theory, where the traditional Feynman diagram expansion hides the true simplicity of the on-shell answer \footnote{See, {\it e.g.}, \cite{Elvang:2015rqa} for a recent review.}.
Moreover, it is our belief that  holographic $n$-point functions of arbitrary KK modes must be completely fixed by general consistency requirements such as superconfomal symmetry and crossing --  this is
a restatement of uniqueness of  the two-derivative action of $10d$ IIB supergravity  (up to field redefinitions). It must then be possible to bypass the diagrammatic expansion altogether and directly  {\it bootstrap} the holographic correlators. The natural language for such an approach is the Mellin representation of conformal field theory (CFT) correlators, initiated by Mack \cite{Mack:2009mi} and developed in \cite{Penedones:2010ue, Fitzpatrick:2011ia, Paulos:2011ie, Fitzpatrick:2011hu, Costa:2012cb, Goncalves:2014rfa}. 
In Mellin space, tree-level $AdS_5$ correlators are rational functions of Mandelstam-like invariants, with poles and residues controlled by factorization, in direct analogy with tree-level scattering  amplitudes in flat space.

In this letter we report an elegant formula for the four-point function of arbitrary single-trace one-half BPS operators in the supergravity limit. 
We have discovered a simple expression 
 that satisfies all consistency conditions and reproduces  all  explicitly calculated examples \cite{Arutyunov:2000py,Arutyunov:2002fh,Arutyunov:2003ae,Uruchurtu:2008kp,Uruchurtu:2011wh}.
 We believe that this is the unique solution of our bootstrap problem, 
 but  a complete proof of uniqueness is presently lacking.

\medskip

\noindent
{\bf Superconformal symmetry.} Let us first review the constraints of superconformal invariance. 
 We focus on  one-half BPS local operators,  
 $\mathcal{O}_{p}^{I_1\ldots I_{p} }(x)= {\rm Tr}\, X^{\{I_1}\ldots X^{I_{p}\} }(x)$, $I_k = 1, \dots 6$, 
  in the symmetric-traceless  representation of the $SO(6)$  
 R-symmetry. It is convenient to keep track of the R-symmetry structure by contracting the $SO(6)$ indices with a null vector,
 \begin{equation} \label{t}
 {O}_{p}(x, t) =  t_{I_1} \dots t_{I_p} \,  {O}_{p}^{I_1\ldots I_{p} } (x) \, , \quad t \cdot t =0\,.
 \end{equation}
 The four-point correlator
 \begin{equation}
 G_{p_1p_2 p_3 p_4} = \langle   {O}_{p_1}   {O}_{p_2}   {O}_{p_3}   {O}_{p_4}\rangle 
 \end{equation}
  is then a function of the four spacetime coordinates $x_i$ and of the four ``internal'' coordinates $t_i$.
Invariance under the conformal group $SO(4, 2)$ and R-symmetry group $SO(6)$ 
implies that it is really a  function  of conformal cross ratios $U$ and $V$ and of R-symmetry cross rations $\sigma$ and $\tau$, up to a kinematic prefactor \footnote{To avoid cluttering, we will henceforth  omit
the labels $p_i$ of the external operators.}:
\begin{equation}
G(x_i,t_i)= \prod_{i<j} \left( \frac{t_{ij}}{x_{ij}^2}\right)^{\gamma^0_{ij}}\left(\frac{t_{12}t_{34}}{x^2_{12}x^2_{34}}\right)^L \mathcal{G}(U,V;\sigma,\tau)\, ,
\end{equation}
where $x_{ij}  = x_i-x_j$, $t_{ij} = t_i  \cdot t_j$ and
\begin{eqnarray}
\nonumber {}&  U  =  \frac{(x_{12})^2(x_{34})^2}{(x_{13})^2(x_{24})^2},\;\;\;\;\;V = \frac{(x_{14})^2(x_{23})^2}{(x_{13})^2(x_{24})^2}\\
 {}&  \sigma = \frac{ t_{13} t_{24} }{ t_{1 2}  t_{34} },\;\;\;\;\;\tau =   \frac{ t_{14} t_{23} }{ t_{1 2}  t_{34} }  \,.
\end{eqnarray}
The exponents $\gamma_{ij}^0$ are given by
\begin{eqnarray}
\nonumber{}&\gamma^0_{12}= \frac{p_1+p_2-p_3-p_4}{2},\;\;\;\;\gamma^0_{13}= \frac{p_1+p_3-p_2-p_4}{2}\\
\nonumber{}&\gamma^0_{34}=\gamma^0_{24}=0,\;\;\;\; \gamma^0_{14}=p_4-L\\
{}&\gamma^0_{23}=p_4-L-\frac{p_1+p_4-p_2-p_3}{2}\,.
\end{eqnarray}
Finally, the exponent $L$ is defined as follows.
Assuming without loss of generality $p_1\geq p_2\geq p_3\geq p_4$, we distinguish two cases: $p_1+p_4\leq p_2+p_3$ (case I) and $p_1+p_4> p_2+p_3$ (case II). Then \footnote{In case II, $L$ is an integer thanks
to $SO(6)$ selection rules.}
\begin{eqnarray}
&L= p_4 \;\;\;\;\;\;\;\; \;\;\;\;\;\;\;\; \;\;\;\;\;\;\; \mathrm{(case\;\; I)}\\
{}&L= \frac{p_2+p_3+p_4-p_1}{2} \;\;\;\;\;\;\;\mathrm{(case\;\; II)}\,.  \nonumber{}
\end{eqnarray}
It immediately follows from these definitions that $\mathcal{G}(U,V;\sigma,\tau)$ is a degree $L$ polynomial in $\sigma$ and $\tau$, 
\begin{equation}
\mathcal{G}(U,V;\sigma,\tau) = \sum_{0 \leq m+n \leq  L} \sigma^m \tau^n  {\cal G}^{(m, n)}(U, V) \,.
\end{equation}
Invariance under the full superconformal symmetry $PSU(2, 2 |4)$ further implies the  Ward identity \cite{Eden:2000bk, Nirschl:2004pa}
\begin{equation}\label{scfwi}
\partial_{\bar{z}}[\mathcal{G}(z\bar{z},(1-z)(1-\bar{z});\alpha\bar{\alpha},(1-\alpha)(1-\bar{\alpha}))\big|_{\bar{\alpha}\to1/\bar{z}}]=0\, ,
\end{equation} 
where we have performed the useful change of variables  $U = z \bar z$, $V = (1-z) (1-\bar z)$, $\sigma = \alpha \bar \alpha$, $\tau = (1-\alpha)(1- \bar \alpha)$.
Its solution can be written as \cite{Eden:2000bk, Nirschl:2004pa}
\begin{equation}\label{split} 
\mathcal{G}(U,V;\sigma,\tau)=\mathcal{G}_{\rm free}(U,V;\sigma,\tau)+
R\,\mathcal{H}(U,V;\sigma,\tau) \, ,
\end{equation}
where $\mathcal{G}_{\rm free}$ is the 
answer in free SYM theory and
\begin{eqnarray}\label{rfactor}
 R & =&\tau \, 1+(1-\sigma-\tau)\, V+(-\tau-\sigma\tau+\tau^2)\, U \\
&&+(\sigma^2-\sigma-\sigma\tau)\, UV+ \sigma V^2+\sigma\tau \, U^2  \nonumber \\
&=&(1-z\alpha)(1-\bar{z}\alpha)(1-z\bar{\alpha})(1-\bar{z}\bar{\alpha})\,.\nonumber
\end{eqnarray}
All dynamical information is contained in the
 {\it a priori} unknown function $\mathcal{H}(U,V;\sigma,\tau)$. 
\medskip

\noindent
{\bf Mellin.}
The  Mellin amplitude ${\cal M}$  is defined as~\cite{Mack:2009mi}
\begin{equation}  \label{remove}
\mathcal{M}(s,t;\sigma,\tau)=\frac{M(s,t;\sigma,\tau)}{\Gamma_{p_1p_2p_3p_4}}\, ,
\end{equation} %
where 
\begin{eqnarray} \label{transform}
\nonumber {}& M(s,t;\sigma,\tau)=\int_0^\infty dV V^{-\frac{t}{2}+\frac{\min\{p_1+p_4,p_2+p_3\}}{2}-1}\\
{}&\int_0^\infty dU U^{-\frac{s}{2}+\frac{p_3+p_4}{2}-L-1}\; \mathcal{G}_{\rm conn}(U,V;\sigma,\tau)\, 
\end{eqnarray}
is an integral transform of the {\it connected} 
    four-point function
   with respect to the conformal cross-ratios, and 
\begin{eqnarray}  \label{Gamma}
 \Gamma_{p_1p_2p_3p_4}=&{}\Gamma[\frac{p_1+p_2-s}{2}]\Gamma[\frac{p_3+p_4-s}{2}]\Gamma[\frac{p_2+p_3-t}{2}]\\
&\; \;\,\Gamma[\frac{p_1+p_4-t}{2}]\Gamma[\frac{p_1+p_3-u}{2}]\Gamma[\frac{p_2+p_4-u}{2}]\, .\nonumber
\end{eqnarray}
We have also defined 
\begin{equation} \label{u}
u=p_1+p_2+p_3+p_4-s-t \, .
\end{equation}
Mack~\cite{Mack:2009mi} observed that ${\cal M}$ behaves in some ways as an S-matrix, with the dual variables $s$, $t$, $u$ playing the role of Mandelstam invariants. 
Its analytic structure is very simple: for fixed $t$,  the so-called reduced  Mellin amplitude ${M}$ has simple poles in $s$; each pole corresponds to an intermediate operator exchanged in the $s$-channel  OPE  of the  four-point function. Organizing operators in conformal families,
 each exchanged primary of dimension $\Delta$ and spin $J$ contributes an infinite sequence of pole at $s = \tau + 2m$, where $\tau = \Delta-J$ is  the twist and $m \in \mathbb{Z}_+$.  
 Analogous statements hold in the crossed channels.

As pointed out by Penedones \cite{Penedones:2010ue}, definition  (\ref{remove})
is completely natural in a large $N$ theory: dividing by $\Gamma_{p_1 p_2 p_3 p_4}$ 
 removes   the poles associated with   double-trace operators, leaving in ${\cal M}$ only  single-trace poles. Recall that  ${\cal G}_{\rm conn}$ is subleading at large $N$ with respect to the disconnected part -- 
it is  $O(1/N^2)$ in $SU(N)$ SYM theory. It 
receives contributions from both from single-trace operators
 and double-trace operators. For example, in the $s$-channel OPE $(x_{12} \, , x_{34} \to 0)$  there are double-trace operators of the schematic form ${\cal O}_{p_1} \partial^J \Box^n {\cal O}_{p_2}$,
of twist $\tau = p_1 + p_2  + 2n + O(1/N^2)$, and ${\cal O}_{p_3} \partial^J \Box^n {\cal O}_{p_4}$, of twist $\tau = p_3 + p_4  + 2n + O(1/N^2)$. Their contribution is precisely captured by the first two Gamma functions
in (\ref{Gamma}), while the other Gamma functions serve the same purpose in the $t$- and $u$-channels \footnote{What's more, since $p_1 + p_2$ and $p_3 + p_4$ differ by an even integer,
 (\ref{Gamma}) has {\it double} poles at $s = p_1 + p_2 + 2 q$, with $q \in \mathbb{Z}_+$.  These  double poles are  needed to produce terms proportional to $\log U$ in ${\cal G}_{\rm conn}$, associated to the $O(1/N^2)$ anomalous dimensions of the double-trace operators. }. 

We are interested in further taking the  't Hooft coupling $\lambda$ to infinity. This is the regime is described in the bulk by classical supergravity.
The only single-trace operators that survive in this limit are one-half BPS operators and their superconformal descendants, dual to supergravity KK modes.
 Naively, each single-trace operator  ${\cal O}$ appearing, {\it e.g.}, in the $s$-channel OPE
 would contribute infinitely many  poles to ${\cal M}$ at $s = \tau_{\cal O} + 2m$, $m \in \mathbb{Z}_+$, but in fact this sequence of  single-trace poles {\it truncates} before it would start overlapping with 
the double-trace poles in (\ref{Gamma}) \footnote{Note that $\tau_{\cal O}$ is an integer with the same parity as $p_1 + p_2$ and $p_3 + p_4$, see Table 1 of \cite{Dolan:2002zh}.}.
This truncation is necessary for a consistent OPE interpretation 
\footnote{For example, a  {\it triple} pole in $M$  would translate
into a $(\log  U)^2$ term in ${\cal G}_{\rm conn}$, which cannot appear at order $O(1/N^2)$. A detailed analysis will be presented in \cite{longpaper}.}.

The same conclusion can be reached by a diagrammatic argument in supergravity. The $O(1/N^2)$ term of ${\cal G}_{\rm conn}$ is given by a (finite) sum of  tree level Witten diagrams: $s$-, $t$- and $u$-channel exchange diagrams, in correspondence with the  single-trace operators exchanged in the respective channel OPE;
 and additional contact diagrams, arising from quartic vertices.
The Mellin amplitude for an  $s$-channel exchange Witten diagram  takes the form \cite{Costa:2012cb}
\begin{equation} \label{witten}
\mathcal{M}_{\Delta, J}(s,t) = 
\sum_{m=0}^{\infty}\frac{Q_{J,m}(t)}{s- (\Delta - J)-2m}+P_{J-1}(s,t)\, ,
\end{equation}
where $\Delta$ and $J$ are the dimension and spin of the exchanged field,  $Q_{J,m}(t)$ are polynomials in $t$ of degree $J$ and $P_{J-1}(s,t)$ is a polynomial in $s$ and $t$ of degree $J-1$.
For the values of $\Delta$ and $J$ that appear in $AdS_5 \times S^5$ supergravity,  the sum over $m$ truncates, 
with the same $m_{\rm max}$ as predicted by the above OPE argument  \footnote{This truncation can be seen from the explicit expressions in \cite{Costa:2012cb} and is equivalent to the observation in \cite{DHoker:1999aa} that the exchange Witten diagrams relevant for $AdS_5 \times S^5$ supergravity can always be written as finite sums of contact diagrams.}.

We see from (\ref{witten}) that exchange diagrams grow at most linearly  at large $s$ and $t$, because $J \leq 2$ in supergravity. 
In Mellin space, a contact diagram is a polynomial in $s$ and $t$  \cite{Penedones:2010ue}, of degree equal to half  the number of spacetime derivatives in the quartic vertex.
The $AdS_5$ effective action \cite{Arutyunov:1999fb} contains quartic vertices with up to four spacetime derivatives, 
which would naively give a quadratic asymptotic growth for large $s$ and $t$, but in fact the final answer is expected to grow at most linearly \footnote{Happily, in all explicit supergravity calculations performed so far \cite{Arutyunov:2000py, Arutyunov:2002fh, Arutyunov:2003ae, Berdichevsky:2007xd,Uruchurtu:2008kp, Uruchurtu:2011wh}
the contribution of  vertices with four spacetime derivatives  can be effectively re-written in terms of vertices with zero or two derivatives. It would be nice to prove  that this reduction of order occurs in the general case,
as we have argued indirectly from compatibility with the flat space limit.
 We are indebted to G. Arutyunov and S. Frolov for a very useful discussion on this point.}.  Indeed, a larger asymptotic growth would be inconsistent with the flat-space space limit~\cite{Penedones:2010ue}.

\medskip
\noindent {\bf A bootstrap problem.}
We are ready to enumerate several  properties of ${\cal M}$.  First, there are 
 structural algebraic properties, valid for any $N$ and $\lambda$:

\noindent
{ 1.} \textit{{Bose symmetry.}} $\mathcal{M}$ is
 invariant under permutation of the Mandelstam variables, if the quantum numbers of the external operators  are permuted accordingly.
For  example, for equal weights $p_i = p$, this gives the usual crossing relations
\begin{equation}
\begin{split}
\sigma^{p} {\mathcal{M}}( u,t;1/\sigma,\tau/\sigma)={}& {\mathcal{M}}(s,t,;\sigma,\tau)\\
\tau^{p} {\mathcal{M}}(t,s;\sigma/\tau,1/\tau)={}& {\mathcal{M}}(s,t;\sigma,\tau)\,,
\end{split}
\end{equation}
where $u$ was defined in (\ref{u}).

\noindent
{2.} \textit{{Superconformal Ward identity.}} We need to translate (\ref{split}) into Mellin space.  In parallel with (\ref{transform}), 
we  take the integral transform
of the dynamical ${\cal H}$ function,
\begin{eqnarray} \label{transformH}
\nonumber {}& \widetilde M (s,t;\sigma,\tau)=\int_0^\infty dV V^{-\frac{t}{2}+\frac{\min\{p_1+p_4,p_2+p_3\}}{2}-1}\\
{}&\int_0^\infty dU U^{-\frac{s}{2}+\frac{p_3+p_4}{2}-L-1}\; \mathcal{H} (U,V;\sigma,\tau)\, 
\end{eqnarray}
and  then define 
\begin{equation}
\widetilde {\cal M} (s, t; \sigma, \tau) = \frac{\widetilde M (s, t; \sigma, \tau)}{\widetilde \Gamma_{p_1 p_2 p_3 p_4}}\, ,
\end{equation}
where $\widetilde \Gamma_{p_1 p_2 p_3 p_4}$ is obtained by replacing $u \to \tilde u = u-4$ in  (\ref{Gamma}). This shift in $u$ is useful
to make the crossing symmetry properties of $\widetilde {\mathcal M}$ more transparent. For example, for equal weights,
\begin{equation}
\begin{split}
\sigma^{p-2} {\widetilde {\mathcal M}}(\tilde u,t;1/\sigma,\tau/\sigma)={}& {  \widetilde {\mathcal M}}(s,t,;\sigma,\tau)\\
\tau^{p-2} {  \widetilde {\mathcal M}}(t,s;\sigma/\tau,1/\tau)={}& { \widetilde {\mathcal M}}(s,t;\sigma,\tau)\,.
\end{split}
\end{equation}
With this definition of $\widetilde {\cal M}$, 
(\ref{split}) is equivalent to
\begin{equation} \label{splitmellin}
\mathcal{M}(s,t;\sigma,\tau)=
\hat{R}\circ \widetilde{\mathcal{M}}(s,t;\sigma,\tau)\, ,
\end{equation}
where $\hat{R}$ is given by (\ref{rfactor}) with each $U^mV^n$ replaced by a difference operator $\widehat{U^mV^n}$ acting 
as
\begin{equation} 
\begin{split}
{}&\widehat{U^mV^n}\circ \widetilde{\mathcal{M}}(s,t;\sigma,\tau) = 
  \widetilde{\mathcal{M}}(s-2m,t-2n;\sigma,\tau)\; \times \\
&\left(\frac{p_1+p_2-s}{2}\right)_m   \left(\frac{p_1+p_3-u}{2}\right)_{2-m-n}   \left(\frac{p_1+p_4-t}{2}\right)_n \nonumber
\\
& \left(\frac{p_2+p_3-t}{2}\right)_n \left(\frac{p_2+p_4-u}{2}\right)_{2-m-n} \left(\frac{p_3+p_4-s}{2}\right)_m \, ,  \nonumber
\end{split}
\end{equation}
with $(h)_n = \frac{\Gamma[h + n]} {\Gamma[h]}$ denoting the Pochhammer symbol. 
Contrasting (\ref{split}) and (\ref{splitmellin}), it may appear that we have forgotten the term ${\cal G}_{\rm free}$. 
In fact,  the Mellin transform of the free part is ``zero'' (a sum of delta functions) and can be consistently ignored. While the  direct Mellin transform (\ref{transform}) 
is unambiguous,
the  inverse Mellin transform from ${\cal M}$ back to ${\cal G}_{\rm conn}$ requires to prescribe an integration contour  -- one must integrate inside the ``fundamental strips'' for $s$ and $t$ 
where the integrals 
in (\ref{transform}) converge. The correct choice of contour reproduces ${\cal G}_{\rm free}$ automatically. Details will appear in \cite{longpaper}.

Second, we have argued that at leading $O(1/N^2)$ order and for $\lambda \to \infty$,  ${\cal M}$ becomes a very constrained rational function:

\noindent
{3.} \textit{{Analytic structure.}}  $\mathcal{M}$   has a finite number of simple poles in $s$, $t$, $u$, at the locations
\begin{eqnarray}
\nonumber s_0 &= & s_M-2 a \, ,  \quad s_0 \geq  2\\
\nonumber t_0 &= & t_M -2 b \, ,  \, \,\quad t_0 \geq 2\\
\nonumber u_0 &= &  u_M - 2 c \, ,  \quad u_0 \geq 2\, ,
\end{eqnarray}
where
\begin{eqnarray}
{}&s_M=\min\{p_1+p_2,p_3+p_4\}-2\\  
{}&t_M=\min\{p_1+p_4,p_2+p_3\}-2\\
{}&{u}_M=\min\{p_1+p_3,p_2+p_4\}-2
\end{eqnarray}
and $a$, $b$, $c$ are non-negative integers. Furthermore,
the residue at each pole is a polynomial in the other Mandelstam variable.

\noindent
{4.}  \textit{{Asymptotics.}}
${\cal M}$ grows linearly at large values of the Mandelstam variables,  \begin{equation}
\label{scaling}
{\mathcal{M}}(\beta s,\beta t;\sigma,\tau) \sim O(\beta) \, \quad {\rm for}\; \beta \to \infty \,.
\end{equation}

Taken together, these conditions define a very constrained bootstrap problem.

\smallskip 

\noindent
{\bf Our solution.}
Some experimentation at low KK levels leads us  to the ansatz
\begin{equation} \label{ansatz}
 \widetilde{\mathcal{M}}(s,t ;\sigma,\tau)= \qquad\qquad \qquad \qquad \qquad \qquad \qquad
 \end{equation}
\vspace{-0.9cm}
 \begin{equation}
 \sum_{{\tiny\begin{split}{}&i+j+k=L-2\\{}&0\leq i,j,k\leq L-2\end{split}}} \frac{a_{ijk}\;\sigma^i\tau^j}{(s-s_M+2k)(t-t_M+2j)(\tilde{u}- {u}_M+2i)}\, . \qquad\nonumber
\end{equation}
This is the most symmetric expression compatible with Bose symmetry, 
 the scaling (\ref{scaling}) and  the expected pole structure. Imposing that  ${\cal M} = \hat R \circ \widetilde {\cal M}$ has poles with polynomial residues 
fixes the coefficients $a_{ijk}$ uniquely, up to overall normalization:
\begin{eqnarray}\label{aijk}
\nonumber a_{ijk}{}&=(1+\frac{|p_1-p_2+p_3-p_4|}{2})_i^{-1}(1+\frac{|p_1+p_4-p_2-p_3|}{2})_j^{-1}\\
{}&\times(1+\frac{|p_1+p_2-p_3-p_4|}{2})_k^{-1} {L-2 \choose i \; j \;k}  \, C_{p_1 p_2 p_3 p_4} \, ,
\end{eqnarray}
where ${L-2 \choose i \; j \;k}$ is the trinomial coefficient. The normalization constant $C_{p_1 p_2 p_3 p_4} = f(p_1, p_2, p_3, p_4)/N^2$ cannot be determined from our homogeneous
consistency conditions \footnote{The normalization can in principle be found by extracting from (\ref{ansatz}) the OPE coefficients of the intermediate one-half BPS operators. Since these OPE coefficients
are protected, they can be
 matched with their free field expressions, thereby fixing the overall constant.}. We have checked that our proposal reproduces all the available supergravity calculations:
the equal weights cases $p_i =2$ \cite{Arutyunov:2000py}, $p_i =3$ \cite{Arutyunov:2002fh} and $p_i = 4$ \cite{Arutyunov:2003ae}, as well as the general expression \cite{Berdichevsky:2007xd,Uruchurtu:2008kp, Uruchurtu:2011wh} for next-to-next extremal correlators ({\it i.e.}, the cases  $p_1 = n+k$, $p_2 = n-k$, $p_3 = p_4 = k+2$).   We have not yet been able to prove, but find it very plausible,
that  (\ref{ansatz}) is the most general ansatz compatible with the bootstrap conditions.

\medskip

\noindent
{\bf A position space method.}
The power of maximal supersymmetry can also be appreciated by an independent method in position space, which will be fully illustrated in \cite{longpaper}. This method 
mimics the
conventional holographic calculation of correlation functions, writing the answer as a sum of exchange and contact Witten diagrams,
\begin{equation}
\mathcal{A}_{\rm sugra}=\mathcal{A}_{\rm exchange}+\mathcal{A}_{\rm contact} \, ,
\end{equation}
but it eschews knowledge of the precise cubic and quartic couplings, left as undetermined coefficients. 
Using the results of  \cite{DHoker:1999aa}, the exchange diagrams are expressed as finite sums of contact diagrams ($\bar{D}$-functions). All in all, one is led to an ansatz in terms a finite sum of $\bar D$-functions,
depending linearly on a set of coefficients, to be fixed by imposing the superconformal Ward identity.
The task of obtaining the correct vertices from the effective action and working out tedious combinatorics is replaced 
by an easier linear algebra problem. 
In practice, one uses the fact that  $\bar{D}$-functions can be uniquely written as
\begin{equation}
\bar{D}_{\Delta_1\Delta_2\Delta_3\Delta_4}=R_\Phi \Phi(U,V)+R_{V} \log V+R_{U} \log U+R_0
\end{equation}
where $\Phi(U,V)=\bar{D}_{1111}$ is the scalar box diagram, and  $R_{\Phi,  U, V,0}$ are rational functions of the cross-ratios $U$ and $V$. 
The ansatz for $\mathcal{A}_{\rm sugra}$ can be decomposed similarly, with rational coefficient functions $R^{\rm sugra} (z, \bar z; \alpha, \bar \alpha)$ that also depend on the R-symmetry cross-ratios.
The superconformal Ward identity  then becomes a set of conditions  on the rational coefficient functions
\begin{equation}
\begin{split}
R_\Phi^{\rm sugra}(z,\bar{z};\alpha,1/\bar{z}){}&=0\\
R_{V}^{\rm sugra}(z,\bar{z};\alpha,1/\bar{z}){}&=0\\
R_{U}^{\rm sugra}(z,\bar{z};\alpha,1/\bar{z}){}&=0\, ,
\end{split}
\end{equation}
giving  a set of linear equations for the undetermined coefficients. Uniqueness of the maximally supersymmetric action guarantees  the existence of a unique solution up to overall rescaling.
Finally, the overall normalization is determined by matching the protected part of the correlator with free field theory,
\begin{equation}
R_{0}^{\rm sugra}(z,\bar{z};\alpha,1/\bar{z}){}= {\cal G}_{\rm free} (z, \bar z; \alpha, 1/\bar z) \,.
\end{equation}

This method is fully rigorous, relying entirely on the structure of the supergravity calculation with no additional assumption. Despite being much simpler than the conventional approach,
even this method quickly becomes unwieldy as the KK level is increased.
We have so far obtained results for the equal weights correlators with
 $p=2,3,4, 5$. The result for $p=5$ is new. It agrees both with our Mellin formula (\ref{ansatz}) 
 and with a previous conjecture
by Dolan, Nirschl and Osborn \cite{Dolan:2006ec}, who 
proposed a general answer for arbitrary equal weights, as a sum of $\bar D$-functions. 
Unfortunately the complexity of their expression grows very rapidly with $p$,
making a check against (\ref{ansatz}) very cumbersome for $p > 5$.

\medskip

\noindent
{\bf Discussion.} 
The remarkable simplicity of the general formula (\ref{ansatz}) is a welcome surprise. Like the Parke-Taylor formula \cite{Parke:1986gb}  for tree-level MHV 
gluon scattering amplitudes, it encodes in a succinct expression the sum of an intimidating number of diagrams. It appears that
holographic correlators  are much simpler than previously understood. We believe that they should be studied following the blueprint of the 
modern on-shell approach to perturbative gauge theory amplitudes. 
While we have obtained (\ref{ansatz}) as the solution of a set of bootstrap conditions, a more constructive approach based on on-shell recursion relations
({\it \`a la} BCFW \cite{Britto:2005fq}?) may also exist, and lend itself more easily to the generalization to higher $n$-point correlators \footnote{A BCFW-inspired formalism for holographic correlators has  been developed
in momentum space \cite{Raju:2010by, Raju:2012zr}.
}.

An important direction to pursue is the generalization of our results to include the 't Hooft coupling dependence. For large $\lambda$, one can study
$\alpha'$ corrections by  relaxing the asymptotic behavior (\ref{scaling}). It would be interesting to make contact with the results of \cite{Alday:2016htq}.
 In the opposite limit of small $\lambda$, it would be worthwhile to explore whether a pattern similar to (\ref{ansatz}) can be recognized in the Mellin transformation of perturbative correlators \footnote{Results for general weights $p_i$ are available up to order $O(\lambda^3)$ \cite{Chicherin:2015edu}.}. 
On a more practical note, (\ref{ansatz})  implicitly contains a large amount of CFT data, such as  the order $O(1/N^2)$ anomalous dimensions of arbitrary double-trace operators in the strong coupling limit. These are useful data
for comparison with the superconformal bootstrap \cite{Beem:2013qxa, Alday:2014qfa}, and it will be nice to extract them explicitly.

Finally, a direct generalization of the approach pursued here gives structurally similar results for holographic correlators in $AdS_7 \times S^4$, as we shall report elsewhere.

\bigskip

\begin{acknowledgments}
Our work is  supported in part by NSF Grant PHY-1316617.
We are grateful to Gleb Arutyunov, Sergey Frolov, Carlo Meneghelli, Jo\~ao Penenones, and Volker Schomerus  for useful conversations.
\end{acknowledgments}

\bibliography{refletter}

\begin{thebibliography}{49}%
\makeatletter
\providecommand \@ifxundefined [1]{%
 \@ifx{#1\undefined}
}%
\providecommand \@ifnum [1]{%
 \ifnum #1\expandafter \@firstoftwo
 \else \expandafter \@secondoftwo
 \fi
}%
\providecommand \@ifx [1]{%
 \ifx #1\expandafter \@firstoftwo
 \else \expandafter \@secondoftwo
 \fi
}%
\providecommand \natexlab [1]{#1}%
\providecommand \enquote  [1]{``#1''}%
\providecommand \bibnamefont  [1]{#1}%
\providecommand \bibfnamefont [1]{#1}%
\providecommand \citenamefont [1]{#1}%
\providecommand \href@noop [0]{\@secondoftwo}%
\providecommand \href [0]{\begingroup \@sanitize@url \@href}%
\providecommand \@href[1]{\@@startlink{#1}\@@href}%
\providecommand \@@href[1]{\endgroup#1\@@endlink}%
\providecommand \@sanitize@url [0]{\catcode `\\12\catcode `\$12\catcode
  `\&12\catcode `\#12\catcode `\^12\catcode `\_12\catcode `\%12\relax}%
\providecommand \@@startlink[1]{}%
\providecommand \@@endlink[0]{}%
\providecommand \url  [0]{\begingroup\@sanitize@url \@url }%
\providecommand \@url [1]{\endgroup\@href {#1}{\urlprefix }}%
\providecommand \urlprefix  [0]{URL }%
\providecommand \Eprint [0]{\href }%
\providecommand \doibase [0]{http://dx.doi.org/}%
\providecommand \selectlanguage [0]{\@gobble}%
\providecommand \bibinfo  [0]{\@secondoftwo}%
\providecommand \bibfield  [0]{\@secondoftwo}%
\providecommand \translation [1]{[#1]}%
\providecommand \BibitemOpen [0]{}%
\providecommand \bibitemStop [0]{}%
\providecommand \bibitemNoStop [0]{.\EOS\space}%
\providecommand \EOS [0]{\spacefactor3000\relax}%
\providecommand \BibitemShut  [1]{\csname bibitem#1\endcsname}%
\let\auto@bib@innerbib\@empty
\bibitem [{\citenamefont {Maldacena}(1999)}]{Maldacena:1997re}%
  \BibitemOpen
  \bibfield  {author} {\bibinfo {author} {\bibfnamefont {J.~M.}\ \bibnamefont
  {Maldacena}},\ }\href {\doibase 10.1023/A:1026654312961} {\bibfield
  {journal} {\bibinfo  {journal} {Int. J. Theor. Phys.}\ }\textbf {\bibinfo
  {volume} {38}},\ \bibinfo {pages} {1113} (\bibinfo {year} {1999})},\ \bibinfo
  {note} {[Adv.~Theor.~Math.~Phys.2,231(1998)]},\ \Eprint
  {http://arxiv.org/abs/hep-th/9711200} {arXiv:hep-th/9711200 [hep-th]}
  \BibitemShut {NoStop}%
\bibitem [{\citenamefont {Gubser}\ \emph {et~al.}(1998)\citenamefont {Gubser},
  \citenamefont {Klebanov},\ and\ \citenamefont {Polyakov}}]{Gubser:1998bc}%
  \BibitemOpen
  \bibfield  {author} {\bibinfo {author} {\bibfnamefont {S.~S.}\ \bibnamefont
  {Gubser}}, \bibinfo {author} {\bibfnamefont {I.~R.}\ \bibnamefont
  {Klebanov}}, \ and\ \bibinfo {author} {\bibfnamefont {A.~M.}\ \bibnamefont
  {Polyakov}},\ }\href {\doibase 10.1016/S0370-2693(98)00377-3} {\bibfield
  {journal} {\bibinfo  {journal} {Phys. Lett.}\ }\textbf {\bibinfo {volume}
  {B428}},\ \bibinfo {pages} {105} (\bibinfo {year} {1998})},\ \Eprint
  {http://arxiv.org/abs/hep-th/9802109} {arXiv:hep-th/9802109 [hep-th]}
  \BibitemShut {NoStop}%
\bibitem [{\citenamefont {Witten}(1998)}]{Witten:1998qj}%
  \BibitemOpen
  \bibfield  {author} {\bibinfo {author} {\bibfnamefont {E.}~\bibnamefont
  {Witten}},\ }\href@noop {} {\bibfield  {journal} {\bibinfo  {journal} {Adv.
  Theor. Math. Phys.}\ }\textbf {\bibinfo {volume} {2}},\ \bibinfo {pages}
  {253} (\bibinfo {year} {1998})},\ \Eprint
  {http://arxiv.org/abs/hep-th/9802150} {arXiv:hep-th/9802150 [hep-th]}
  \BibitemShut {NoStop}%
\bibitem [{Note1()}]{Note1}%
  \BibitemOpen
  \bibinfo {note} {Three-point correlators of arbitrary one-half BPS operators
  were computed in the early days of AdS/CFT \cite {Freedman:1998tz,
  Lee:1998bxa} and found to agree with the tree-level SYM results, a fact
  subsequently explained by a non-renormalization theorem, see, {\protect \it
  e.g.}, \cite {Baggio:2012rr} and references therein.}\BibitemShut {Stop}%
\bibitem [{\citenamefont {Arutyunov}\ and\ \citenamefont
  {Frolov}(2000{\natexlab{a}})}]{Arutyunov:1999fb}%
  \BibitemOpen
  \bibfield  {author} {\bibinfo {author} {\bibfnamefont {G.}~\bibnamefont
  {Arutyunov}}\ and\ \bibinfo {author} {\bibfnamefont {S.}~\bibnamefont
  {Frolov}},\ }\href {\doibase 10.1016/S0550-3213(00)00210-8} {\bibfield
  {journal} {\bibinfo  {journal} {Nucl. Phys.}\ }\textbf {\bibinfo {volume}
  {B579}},\ \bibinfo {pages} {117} (\bibinfo {year} {2000}{\natexlab{a}})},\
  \Eprint {http://arxiv.org/abs/hep-th/9912210} {arXiv:hep-th/9912210 [hep-th]}
  \BibitemShut {NoStop}%
\bibitem [{\citenamefont {D'Hoker}\ \emph
  {et~al.}(1999{\natexlab{a}})\citenamefont {D'Hoker}, \citenamefont
  {Freedman}, \citenamefont {Mathur}, \citenamefont {Matusis},\ and\
  \citenamefont {Rastelli}}]{DHoker:1999pj}%
  \BibitemOpen
  \bibfield  {author} {\bibinfo {author} {\bibfnamefont {E.}~\bibnamefont
  {D'Hoker}}, \bibinfo {author} {\bibfnamefont {D.~Z.}\ \bibnamefont
  {Freedman}}, \bibinfo {author} {\bibfnamefont {S.~D.}\ \bibnamefont
  {Mathur}}, \bibinfo {author} {\bibfnamefont {A.}~\bibnamefont {Matusis}}, \
  and\ \bibinfo {author} {\bibfnamefont {L.}~\bibnamefont {Rastelli}},\ }\href
  {\doibase 10.1016/S0550-3213(99)00525-8} {\bibfield  {journal} {\bibinfo
  {journal} {Nucl. Phys.}\ }\textbf {\bibinfo {volume} {B562}},\ \bibinfo
  {pages} {353} (\bibinfo {year} {1999}{\natexlab{a}})},\ \Eprint
  {http://arxiv.org/abs/hep-th/9903196} {arXiv:hep-th/9903196 [hep-th]}
  \BibitemShut {NoStop}%
\bibitem [{\citenamefont {Arutyunov}\ and\ \citenamefont
  {Frolov}(2000{\natexlab{b}})}]{Arutyunov:2000py}%
  \BibitemOpen
  \bibfield  {author} {\bibinfo {author} {\bibfnamefont {G.}~\bibnamefont
  {Arutyunov}}\ and\ \bibinfo {author} {\bibfnamefont {S.}~\bibnamefont
  {Frolov}},\ }\href {\doibase 10.1103/PhysRevD.62.064016} {\bibfield
  {journal} {\bibinfo  {journal} {Phys. Rev.}\ }\textbf {\bibinfo {volume}
  {D62}},\ \bibinfo {pages} {064016} (\bibinfo {year} {2000}{\natexlab{b}})},\
  \Eprint {http://arxiv.org/abs/hep-th/0002170} {arXiv:hep-th/0002170 [hep-th]}
  \BibitemShut {NoStop}%
\bibitem [{\citenamefont {Dolan}\ and\ \citenamefont
  {Osborn}(2006)}]{Dolan:2004iy}%
  \BibitemOpen
  \bibfield  {author} {\bibinfo {author} {\bibfnamefont {F.~A.}\ \bibnamefont
  {Dolan}}\ and\ \bibinfo {author} {\bibfnamefont {H.}~\bibnamefont {Osborn}},\
  }\href {\doibase 10.1016/j.aop.2005.07.005} {\bibfield  {journal} {\bibinfo
  {journal} {Annals Phys.}\ }\textbf {\bibinfo {volume} {321}},\ \bibinfo
  {pages} {581} (\bibinfo {year} {2006})},\ \Eprint
  {http://arxiv.org/abs/hep-th/0412335} {arXiv:hep-th/0412335 [hep-th]}
  \BibitemShut {NoStop}%
\bibitem [{Note2()}]{Note2}%
  \BibitemOpen
  \bibinfo {note} {See, {\protect \it e.g.}, \cite {Elvang:2015rqa} for a
  recent review.}\BibitemShut {Stop}%
\bibitem [{\citenamefont {Mack}(2009)}]{Mack:2009mi}%
  \BibitemOpen
  \bibfield  {author} {\bibinfo {author} {\bibfnamefont {G.}~\bibnamefont
  {Mack}},\ }\href@noop {} {\  (\bibinfo {year} {2009})},\ \Eprint
  {http://arxiv.org/abs/0907.2407} {arXiv:0907.2407 [hep-th]} \BibitemShut
  {NoStop}%
\bibitem [{\citenamefont {Penedones}(2011)}]{Penedones:2010ue}%
  \BibitemOpen
  \bibfield  {author} {\bibinfo {author} {\bibfnamefont {J.}~\bibnamefont
  {Penedones}},\ }\href {\doibase 10.1007/JHEP03(2011)025} {\bibfield
  {journal} {\bibinfo  {journal} {JHEP}\ }\textbf {\bibinfo {volume} {03}},\
  \bibinfo {pages} {025} (\bibinfo {year} {2011})},\ \Eprint
  {http://arxiv.org/abs/1011.1485} {arXiv:1011.1485 [hep-th]} \BibitemShut
  {NoStop}%
\bibitem [{\citenamefont {Fitzpatrick}\ \emph {et~al.}(2011)\citenamefont
  {Fitzpatrick}, \citenamefont {Kaplan}, \citenamefont {Penedones},
  \citenamefont {Raju},\ and\ \citenamefont {van Rees}}]{Fitzpatrick:2011ia}%
  \BibitemOpen
  \bibfield  {author} {\bibinfo {author} {\bibfnamefont {A.~L.}\ \bibnamefont
  {Fitzpatrick}}, \bibinfo {author} {\bibfnamefont {J.}~\bibnamefont {Kaplan}},
  \bibinfo {author} {\bibfnamefont {J.}~\bibnamefont {Penedones}}, \bibinfo
  {author} {\bibfnamefont {S.}~\bibnamefont {Raju}}, \ and\ \bibinfo {author}
  {\bibfnamefont {B.~C.}\ \bibnamefont {van Rees}},\ }\href {\doibase
  10.1007/JHEP11(2011)095} {\bibfield  {journal} {\bibinfo  {journal} {JHEP}\
  }\textbf {\bibinfo {volume} {11}},\ \bibinfo {pages} {095} (\bibinfo {year}
  {2011})},\ \Eprint {http://arxiv.org/abs/1107.1499} {arXiv:1107.1499
  [hep-th]} \BibitemShut {NoStop}%
\bibitem [{\citenamefont {Paulos}(2011)}]{Paulos:2011ie}%
  \BibitemOpen
  \bibfield  {author} {\bibinfo {author} {\bibfnamefont {M.~F.}\ \bibnamefont
  {Paulos}},\ }\href {\doibase 10.1007/JHEP10(2011)074} {\bibfield  {journal}
  {\bibinfo  {journal} {JHEP}\ }\textbf {\bibinfo {volume} {10}},\ \bibinfo
  {pages} {074} (\bibinfo {year} {2011})},\ \Eprint
  {http://arxiv.org/abs/1107.1504} {arXiv:1107.1504 [hep-th]} \BibitemShut
  {NoStop}%
\bibitem [{\citenamefont {Fitzpatrick}\ and\ \citenamefont
  {Kaplan}(2012)}]{Fitzpatrick:2011hu}%
  \BibitemOpen
  \bibfield  {author} {\bibinfo {author} {\bibfnamefont {A.~L.}\ \bibnamefont
  {Fitzpatrick}}\ and\ \bibinfo {author} {\bibfnamefont {J.}~\bibnamefont
  {Kaplan}},\ }\href {\doibase 10.1007/JHEP10(2012)127} {\bibfield  {journal}
  {\bibinfo  {journal} {JHEP}\ }\textbf {\bibinfo {volume} {10}},\ \bibinfo
  {pages} {127} (\bibinfo {year} {2012})},\ \Eprint
  {http://arxiv.org/abs/1111.6972} {arXiv:1111.6972 [hep-th]} \BibitemShut
  {NoStop}%
\bibitem [{\citenamefont {Costa}\ \emph {et~al.}(2012)\citenamefont {Costa},
  \citenamefont {Goncalves},\ and\ \citenamefont {Penedones}}]{Costa:2012cb}%
  \BibitemOpen
  \bibfield  {author} {\bibinfo {author} {\bibfnamefont {M.~S.}\ \bibnamefont
  {Costa}}, \bibinfo {author} {\bibfnamefont {V.}~\bibnamefont {Goncalves}}, \
  and\ \bibinfo {author} {\bibfnamefont {J.}~\bibnamefont {Penedones}},\ }\href
  {\doibase 10.1007/JHEP12(2012)091} {\bibfield  {journal} {\bibinfo  {journal}
  {JHEP}\ }\textbf {\bibinfo {volume} {12}},\ \bibinfo {pages} {091} (\bibinfo
  {year} {2012})},\ \Eprint {http://arxiv.org/abs/1209.4355} {arXiv:1209.4355
  [hep-th]} \BibitemShut {NoStop}%
\bibitem [{\citenamefont {Gon{\c c}alves}\ \emph {et~al.}(2015)\citenamefont
  {Gon{\c c}alves}, \citenamefont {Penedones},\ and\ \citenamefont
  {Trevisani}}]{Goncalves:2014rfa}%
  \BibitemOpen
  \bibfield  {author} {\bibinfo {author} {\bibfnamefont {V.}~\bibnamefont
  {Gon{\c c}alves}}, \bibinfo {author} {\bibfnamefont {J.}~\bibnamefont
  {Penedones}}, \ and\ \bibinfo {author} {\bibfnamefont {E.}~\bibnamefont
  {Trevisani}},\ }\href {\doibase 10.1007/JHEP10(2015)040} {\bibfield
  {journal} {\bibinfo  {journal} {JHEP}\ }\textbf {\bibinfo {volume} {10}},\
  \bibinfo {pages} {040} (\bibinfo {year} {2015})},\ \Eprint
  {http://arxiv.org/abs/1410.4185} {arXiv:1410.4185 [hep-th]} \BibitemShut
  {NoStop}%
\bibitem [{\citenamefont {Arutyunov}\ \emph {et~al.}(2003)\citenamefont
  {Arutyunov}, \citenamefont {Dolan}, \citenamefont {Osborn},\ and\
  \citenamefont {Sokatchev}}]{Arutyunov:2002fh}%
  \BibitemOpen
  \bibfield  {author} {\bibinfo {author} {\bibfnamefont {G.}~\bibnamefont
  {Arutyunov}}, \bibinfo {author} {\bibfnamefont {F.~A.}\ \bibnamefont
  {Dolan}}, \bibinfo {author} {\bibfnamefont {H.}~\bibnamefont {Osborn}}, \
  and\ \bibinfo {author} {\bibfnamefont {E.}~\bibnamefont {Sokatchev}},\ }\href
  {\doibase 10.1016/S0550-3213(03)00448-6} {\bibfield  {journal} {\bibinfo
  {journal} {Nucl. Phys.}\ }\textbf {\bibinfo {volume} {B665}},\ \bibinfo
  {pages} {273} (\bibinfo {year} {2003})},\ \Eprint
  {http://arxiv.org/abs/hep-th/0212116} {arXiv:hep-th/0212116 [hep-th]}
  \BibitemShut {NoStop}%
\bibitem [{\citenamefont {Arutyunov}\ and\ \citenamefont
  {Sokatchev}(2003)}]{Arutyunov:2003ae}%
  \BibitemOpen
  \bibfield  {author} {\bibinfo {author} {\bibfnamefont {G.}~\bibnamefont
  {Arutyunov}}\ and\ \bibinfo {author} {\bibfnamefont {E.}~\bibnamefont
  {Sokatchev}},\ }\href {\doibase 10.1016/S0550-3213(03)00353-5} {\bibfield
  {journal} {\bibinfo  {journal} {Nucl. Phys.}\ }\textbf {\bibinfo {volume}
  {B663}},\ \bibinfo {pages} {163} (\bibinfo {year} {2003})},\ \Eprint
  {http://arxiv.org/abs/hep-th/0301058} {arXiv:hep-th/0301058 [hep-th]}
  \BibitemShut {NoStop}%
\bibitem [{\citenamefont {Uruchurtu}(2009)}]{Uruchurtu:2008kp}%
  \BibitemOpen
  \bibfield  {author} {\bibinfo {author} {\bibfnamefont {L.~I.}\ \bibnamefont
  {Uruchurtu}},\ }\href {\doibase 10.1088/1126-6708/2009/03/133} {\bibfield
  {journal} {\bibinfo  {journal} {JHEP}\ }\textbf {\bibinfo {volume} {03}},\
  \bibinfo {pages} {133} (\bibinfo {year} {2009})},\ \Eprint
  {http://arxiv.org/abs/0811.2320} {arXiv:0811.2320 [hep-th]} \BibitemShut
  {NoStop}%
\bibitem [{\citenamefont {Uruchurtu}(2011)}]{Uruchurtu:2011wh}%
  \BibitemOpen
  \bibfield  {author} {\bibinfo {author} {\bibfnamefont {L.~I.}\ \bibnamefont
  {Uruchurtu}},\ }\href {\doibase 10.1007/JHEP08(2011)133} {\bibfield
  {journal} {\bibinfo  {journal} {JHEP}\ }\textbf {\bibinfo {volume} {08}},\
  \bibinfo {pages} {133} (\bibinfo {year} {2011})},\ \Eprint
  {http://arxiv.org/abs/1106.0630} {arXiv:1106.0630 [hep-th]} \BibitemShut
  {NoStop}%
\bibitem [{Note3()}]{Note3}%
  \BibitemOpen
  \bibinfo {note} {To avoid cluttering, we will henceforth omit the labels
  $p_i$ of the external operators.}\BibitemShut {Stop}%
\bibitem [{Note4()}]{Note4}%
  \BibitemOpen
  \bibinfo {note} {In case II, $L$ is an integer thanks to $SO(6)$ selection
  rules.}\BibitemShut {Stop}%
\bibitem [{\citenamefont {Eden}\ \emph {et~al.}(2001)\citenamefont {Eden},
  \citenamefont {Petkou}, \citenamefont {Schubert},\ and\ \citenamefont
  {Sokatchev}}]{Eden:2000bk}%
  \BibitemOpen
  \bibfield  {author} {\bibinfo {author} {\bibfnamefont {B.}~\bibnamefont
  {Eden}}, \bibinfo {author} {\bibfnamefont {A.~C.}\ \bibnamefont {Petkou}},
  \bibinfo {author} {\bibfnamefont {C.}~\bibnamefont {Schubert}}, \ and\
  \bibinfo {author} {\bibfnamefont {E.}~\bibnamefont {Sokatchev}},\ }\href
  {\doibase 10.1016/S0550-3213(01)00151-1} {\bibfield  {journal} {\bibinfo
  {journal} {Nucl. Phys.}\ }\textbf {\bibinfo {volume} {B607}},\ \bibinfo
  {pages} {191} (\bibinfo {year} {2001})},\ \Eprint
  {http://arxiv.org/abs/hep-th/0009106} {arXiv:hep-th/0009106 [hep-th]}
  \BibitemShut {NoStop}%
\bibitem [{\citenamefont {Nirschl}\ and\ \citenamefont
  {Osborn}(2005)}]{Nirschl:2004pa}%
  \BibitemOpen
  \bibfield  {author} {\bibinfo {author} {\bibfnamefont {M.}~\bibnamefont
  {Nirschl}}\ and\ \bibinfo {author} {\bibfnamefont {H.}~\bibnamefont
  {Osborn}},\ }\href {\doibase 10.1016/j.nuclphysb.2005.01.013} {\bibfield
  {journal} {\bibinfo  {journal} {Nucl. Phys.}\ }\textbf {\bibinfo {volume}
  {B711}},\ \bibinfo {pages} {409} (\bibinfo {year} {2005})},\ \Eprint
  {http://arxiv.org/abs/hep-th/0407060} {arXiv:hep-th/0407060 [hep-th]}
  \BibitemShut {NoStop}%
\bibitem [{Note5()}]{Note5}%
  \BibitemOpen
  \bibinfo {note} {What's more, since $p_1 + p_2$ and $p_3 + p_4$ differ by an
  even integer, (\ref {Gamma}) has {\protect \it double} poles at $s = p_1 +
  p_2 + 2 q$, with $q \in \protect \mathbb {Z}_+$. These double poles are
  needed to produce terms proportional to $\protect \qopname \relax o{log}U$ in
  ${\protect \cal G}_{\protect \rm conn}$, associated to the $O(1/N^2)$
  anomalous dimensions of the double-trace operators.}\BibitemShut {Stop}%
\bibitem [{Note6()}]{Note6}%
  \BibitemOpen
  \bibinfo {note} {Note that $\tau _{\protect \cal O}$ is an integer with the
  same parity as $p_1 + p_2$ and $p_3 + p_4$, see Table 1 of \cite
  {Dolan:2002zh}.}\BibitemShut {Stop}%
\bibitem [{Note7()}]{Note7}%
  \BibitemOpen
  \bibinfo {note} {For example, a {\protect \it triple} pole in $M$ would
  translate into a $(\protect \qopname \relax o{log}U)^2$ term in ${\protect
  \cal G}_{\protect \rm conn}$, which cannot appear at order $O(1/N^2)$. A
  detailed analysis will be presented in \cite {longpaper}.}\BibitemShut
  {Stop}%
\bibitem [{Note8()}]{Note8}%
  \BibitemOpen
  \bibinfo {note} {This truncation can be seen from the explicit expressions in
  \cite {Costa:2012cb} and is equivalent to the observation in \cite
  {DHoker:1999aa} that the exchange Witten diagrams relevant for $AdS_5 \times
  S^5$ supergravity can always be written as finite sums of contact
  diagrams.}\BibitemShut {Stop}%
\bibitem [{Note9()}]{Note9}%
  \BibitemOpen
  \bibinfo {note} {Happily, in all explicit supergravity calculations performed
  so far \cite {Arutyunov:2000py, Arutyunov:2002fh, Arutyunov:2003ae,
  Berdichevsky:2007xd,Uruchurtu:2008kp, Uruchurtu:2011wh} the contribution of
  vertices with four spacetime derivatives can be effectively re-written in
  terms of vertices with zero or two derivatives. It would be nice to prove
  that this reduction of order occurs in the general case, as we have argued
  indirectly from compatibility with the flat space limit. We are indebted to
  G. Arutyunov and S. Frolov for a very useful discussion on this
  point.}\BibitemShut {Stop}%
\bibitem [{\citenamefont {Rastelli}\ and\ \citenamefont {Zhou}()}]{longpaper}%
  \BibitemOpen
  \bibfield  {author} {\bibinfo {author} {\bibfnamefont {L.}~\bibnamefont
  {Rastelli}}\ and\ \bibinfo {author} {\bibfnamefont {X.}~\bibnamefont
  {Zhou}},\ }\href@noop {} {\bibinfo  {journal} {{to appear}}\ }\BibitemShut
  {NoStop}%
\bibitem [{Note10()}]{Note10}%
  \BibitemOpen
\bibfield  {journal} {  }\bibinfo {note} {The normalization can in principle be
  found by extracting from (\ref {ansatz}) the OPE coefficients of the
  intermediate one-half BPS operators. Since these OPE coefficients are
  protected, they can be matched with their free field expressions, thereby
  fixing the overall constant.}\BibitemShut {Stop}%
\bibitem [{\citenamefont {Berdichevsky}\ and\ \citenamefont
  {Naaijkens}(2008)}]{Berdichevsky:2007xd}%
  \BibitemOpen
  \bibfield  {author} {\bibinfo {author} {\bibfnamefont {L.}~\bibnamefont
  {Berdichevsky}}\ and\ \bibinfo {author} {\bibfnamefont {P.}~\bibnamefont
  {Naaijkens}},\ }\href {\doibase 10.1088/1126-6708/2008/01/071} {\bibfield
  {journal} {\bibinfo  {journal} {JHEP}\ }\textbf {\bibinfo {volume} {01}},\
  \bibinfo {pages} {071} (\bibinfo {year} {2008})},\ \Eprint
  {http://arxiv.org/abs/0709.1365} {arXiv:0709.1365 [hep-th]} \BibitemShut
  {NoStop}%
\bibitem [{\citenamefont {D'Hoker}\ \emph
  {et~al.}(1999{\natexlab{b}})\citenamefont {D'Hoker}, \citenamefont
  {Freedman},\ and\ \citenamefont {Rastelli}}]{DHoker:1999aa}%
  \BibitemOpen
  \bibfield  {author} {\bibinfo {author} {\bibfnamefont {E.}~\bibnamefont
  {D'Hoker}}, \bibinfo {author} {\bibfnamefont {D.~Z.}\ \bibnamefont
  {Freedman}}, \ and\ \bibinfo {author} {\bibfnamefont {L.}~\bibnamefont
  {Rastelli}},\ }\href@noop {} {\bibfield  {journal} {\bibinfo  {journal}
  {Nucl. Phys.}\ }\textbf {\bibinfo {volume} {B562}},\ \bibinfo {pages} {395}
  (\bibinfo {year} {1999}{\natexlab{b}})}\BibitemShut {NoStop}%
\bibitem [{\citenamefont {Dolan}\ \emph {et~al.}(2006)\citenamefont {Dolan},
  \citenamefont {Nirschl},\ and\ \citenamefont {Osborn}}]{Dolan:2006ec}%
  \BibitemOpen
  \bibfield  {author} {\bibinfo {author} {\bibfnamefont {F.~A.}\ \bibnamefont
  {Dolan}}, \bibinfo {author} {\bibfnamefont {M.}~\bibnamefont {Nirschl}}, \
  and\ \bibinfo {author} {\bibfnamefont {H.}~\bibnamefont {Osborn}},\ }\href
  {\doibase 10.1016/j.nuclphysb.2006.05.009} {\bibfield  {journal} {\bibinfo
  {journal} {Nucl. Phys.}\ }\textbf {\bibinfo {volume} {B749}},\ \bibinfo
  {pages} {109} (\bibinfo {year} {2006})},\ \Eprint
  {http://arxiv.org/abs/hep-th/0601148} {arXiv:hep-th/0601148 [hep-th]}
  \BibitemShut {NoStop}%
\bibitem [{\citenamefont {Parke}\ and\ \citenamefont
  {Taylor}(1986)}]{Parke:1986gb}%
  \BibitemOpen
  \bibfield  {author} {\bibinfo {author} {\bibfnamefont {S.~J.}\ \bibnamefont
  {Parke}}\ and\ \bibinfo {author} {\bibfnamefont {T.~R.}\ \bibnamefont
  {Taylor}},\ }\href {\doibase 10.1103/PhysRevLett.56.2459} {\bibfield
  {journal} {\bibinfo  {journal} {Phys. Rev. Lett.}\ }\textbf {\bibinfo
  {volume} {56}},\ \bibinfo {pages} {2459} (\bibinfo {year}
  {1986})}\BibitemShut {NoStop}%
\bibitem [{\citenamefont {Britto}\ \emph {et~al.}(2005)\citenamefont {Britto},
  \citenamefont {Cachazo}, \citenamefont {Feng},\ and\ \citenamefont
  {Witten}}]{Britto:2005fq}%
  \BibitemOpen
  \bibfield  {author} {\bibinfo {author} {\bibfnamefont {R.}~\bibnamefont
  {Britto}}, \bibinfo {author} {\bibfnamefont {F.}~\bibnamefont {Cachazo}},
  \bibinfo {author} {\bibfnamefont {B.}~\bibnamefont {Feng}}, \ and\ \bibinfo
  {author} {\bibfnamefont {E.}~\bibnamefont {Witten}},\ }\href {\doibase
  10.1103/PhysRevLett.94.181602} {\bibfield  {journal} {\bibinfo  {journal}
  {Phys. Rev. Lett.}\ }\textbf {\bibinfo {volume} {94}},\ \bibinfo {pages}
  {181602} (\bibinfo {year} {2005})},\ \Eprint
  {http://arxiv.org/abs/hep-th/0501052} {arXiv:hep-th/0501052 [hep-th]}
  \BibitemShut {NoStop}%
\bibitem [{Note11()}]{Note11}%
  \BibitemOpen
  \bibinfo {note} {A BCFW-inspired formalism for holographic correlators has
  been developed in momentum space \cite {Raju:2010by,
  Raju:2012zr}.}\BibitemShut {Stop}%
\bibitem [{\citenamefont {Alday}\ and\ \citenamefont
  {Bissi}(2016)}]{Alday:2016htq}%
  \BibitemOpen
  \bibfield  {author} {\bibinfo {author} {\bibfnamefont {L.~F.}\ \bibnamefont
  {Alday}}\ and\ \bibinfo {author} {\bibfnamefont {A.}~\bibnamefont {Bissi}},\
  }\href@noop {} {\  (\bibinfo {year} {2016})},\ \Eprint
  {http://arxiv.org/abs/1606.09593} {arXiv:1606.09593 [hep-th]} \BibitemShut
  {NoStop}%
\bibitem [{Note12()}]{Note12}%
  \BibitemOpen
  \bibinfo {note} {Results for general weights $p_i$ are available up to order
  $O(\lambda ^3)$ \cite {Chicherin:2015edu}.}\BibitemShut {Stop}%
\bibitem [{\citenamefont {Beem}\ \emph {et~al.}(2013)\citenamefont {Beem},
  \citenamefont {Rastelli},\ and\ \citenamefont {van Rees}}]{Beem:2013qxa}%
  \BibitemOpen
  \bibfield  {author} {\bibinfo {author} {\bibfnamefont {C.}~\bibnamefont
  {Beem}}, \bibinfo {author} {\bibfnamefont {L.}~\bibnamefont {Rastelli}}, \
  and\ \bibinfo {author} {\bibfnamefont {B.~C.}\ \bibnamefont {van Rees}},\
  }\href {\doibase 10.1103/PhysRevLett.111.071601} {\bibfield  {journal}
  {\bibinfo  {journal} {Phys. Rev. Lett.}\ }\textbf {\bibinfo {volume} {111}},\
  \bibinfo {pages} {071601} (\bibinfo {year} {2013})},\ \Eprint
  {http://arxiv.org/abs/1304.1803} {arXiv:1304.1803 [hep-th]} \BibitemShut
  {NoStop}%
\bibitem [{\citenamefont {Alday}\ and\ \citenamefont
  {Bissi}(2015)}]{Alday:2014qfa}%
  \BibitemOpen
  \bibfield  {author} {\bibinfo {author} {\bibfnamefont {L.~F.}\ \bibnamefont
  {Alday}}\ and\ \bibinfo {author} {\bibfnamefont {A.}~\bibnamefont {Bissi}},\
  }\href {\doibase 10.1007/JHEP02(2015)101} {\bibfield  {journal} {\bibinfo
  {journal} {JHEP}\ }\textbf {\bibinfo {volume} {02}},\ \bibinfo {pages} {101}
  (\bibinfo {year} {2015})},\ \Eprint {http://arxiv.org/abs/1404.5864}
  {arXiv:1404.5864 [hep-th]} \BibitemShut {NoStop}%
\bibitem [{\citenamefont {Freedman}\ \emph {et~al.}(1999)\citenamefont
  {Freedman}, \citenamefont {Mathur}, \citenamefont {Matusis},\ and\
  \citenamefont {Rastelli}}]{Freedman:1998tz}%
  \BibitemOpen
  \bibfield  {author} {\bibinfo {author} {\bibfnamefont {D.~Z.}\ \bibnamefont
  {Freedman}}, \bibinfo {author} {\bibfnamefont {S.~D.}\ \bibnamefont
  {Mathur}}, \bibinfo {author} {\bibfnamefont {A.}~\bibnamefont {Matusis}}, \
  and\ \bibinfo {author} {\bibfnamefont {L.}~\bibnamefont {Rastelli}},\ }\href
  {\doibase 10.1016/S0550-3213(99)00053-X} {\bibfield  {journal} {\bibinfo
  {journal} {Nucl. Phys.}\ }\textbf {\bibinfo {volume} {B546}},\ \bibinfo
  {pages} {96} (\bibinfo {year} {1999})},\ \Eprint
  {http://arxiv.org/abs/hep-th/9804058} {arXiv:hep-th/9804058 [hep-th]}
  \BibitemShut {NoStop}%
\bibitem [{\citenamefont {Lee}\ \emph {et~al.}(1998)\citenamefont {Lee},
  \citenamefont {Minwalla}, \citenamefont {Rangamani},\ and\ \citenamefont
  {Seiberg}}]{Lee:1998bxa}%
  \BibitemOpen
  \bibfield  {author} {\bibinfo {author} {\bibfnamefont {S.}~\bibnamefont
  {Lee}}, \bibinfo {author} {\bibfnamefont {S.}~\bibnamefont {Minwalla}},
  \bibinfo {author} {\bibfnamefont {M.}~\bibnamefont {Rangamani}}, \ and\
  \bibinfo {author} {\bibfnamefont {N.}~\bibnamefont {Seiberg}},\ }\href@noop
  {} {\bibfield  {journal} {\bibinfo  {journal} {Adv. Theor. Math. Phys.}\
  }\textbf {\bibinfo {volume} {2}},\ \bibinfo {pages} {697} (\bibinfo {year}
  {1998})},\ \Eprint {http://arxiv.org/abs/hep-th/9806074}
  {arXiv:hep-th/9806074 [hep-th]} \BibitemShut {NoStop}%
\bibitem [{\citenamefont {Baggio}\ \emph {et~al.}(2012)\citenamefont {Baggio},
  \citenamefont {de~Boer},\ and\ \citenamefont {Papadodimas}}]{Baggio:2012rr}%
  \BibitemOpen
  \bibfield  {author} {\bibinfo {author} {\bibfnamefont {M.}~\bibnamefont
  {Baggio}}, \bibinfo {author} {\bibfnamefont {J.}~\bibnamefont {de~Boer}}, \
  and\ \bibinfo {author} {\bibfnamefont {K.}~\bibnamefont {Papadodimas}},\
  }\href {\doibase 10.1007/JHEP07(2012)137} {\bibfield  {journal} {\bibinfo
  {journal} {JHEP}\ }\textbf {\bibinfo {volume} {07}},\ \bibinfo {pages} {137}
  (\bibinfo {year} {2012})},\ \Eprint {http://arxiv.org/abs/1203.1036}
  {arXiv:1203.1036 [hep-th]} \BibitemShut {NoStop}%
\bibitem [{\citenamefont {Elvang}\ and\ \citenamefont
  {Huang}(2015)}]{Elvang:2015rqa}%
  \BibitemOpen
  \bibfield  {author} {\bibinfo {author} {\bibfnamefont {H.}~\bibnamefont
  {Elvang}}\ and\ \bibinfo {author} {\bibfnamefont {Y.-t.}\ \bibnamefont
  {Huang}},\ }\href
  {http://www.cambridge.org/mw/academic/subjects/physics/theoretical-physics-and-mathematical-physics/scattering-amplitudes-gauge-theory-and-gravity?format=AR}
  {\emph {\bibinfo {title} {{Scattering Amplitudes in Gauge Theory and
  Gravity}}}}\ (\bibinfo  {publisher} {Cambridge University Press},\ \bibinfo
  {year} {2015})\BibitemShut {NoStop}%
\bibitem [{\citenamefont {Dolan}\ and\ \citenamefont
  {Osborn}(2003)}]{Dolan:2002zh}%
  \BibitemOpen
  \bibfield  {author} {\bibinfo {author} {\bibfnamefont {F.~A.}\ \bibnamefont
  {Dolan}}\ and\ \bibinfo {author} {\bibfnamefont {H.}~\bibnamefont {Osborn}},\
  }\href {\doibase 10.1016/S0003-4916(03)00074-5} {\bibfield  {journal}
  {\bibinfo  {journal} {Annals Phys.}\ }\textbf {\bibinfo {volume} {307}},\
  \bibinfo {pages} {41} (\bibinfo {year} {2003})},\ \Eprint
  {http://arxiv.org/abs/hep-th/0209056} {arXiv:hep-th/0209056 [hep-th]}
  \BibitemShut {NoStop}%
\bibitem [{\citenamefont {Raju}(2011)}]{Raju:2010by}%
  \BibitemOpen
  \bibfield  {author} {\bibinfo {author} {\bibfnamefont {S.}~\bibnamefont
  {Raju}},\ }\href {\doibase 10.1103/PhysRevLett.106.091601} {\bibfield
  {journal} {\bibinfo  {journal} {Phys. Rev. Lett.}\ }\textbf {\bibinfo
  {volume} {106}},\ \bibinfo {pages} {091601} (\bibinfo {year} {2011})},\
  \Eprint {http://arxiv.org/abs/1011.0780} {arXiv:1011.0780 [hep-th]}
  \BibitemShut {NoStop}%
\bibitem [{\citenamefont {Raju}(2012)}]{Raju:2012zr}%
  \BibitemOpen
  \bibfield  {author} {\bibinfo {author} {\bibfnamefont {S.}~\bibnamefont
  {Raju}},\ }\href {\doibase 10.1103/PhysRevD.85.126009} {\bibfield  {journal}
  {\bibinfo  {journal} {Phys. Rev.}\ }\textbf {\bibinfo {volume} {D85}},\
  \bibinfo {pages} {126009} (\bibinfo {year} {2012})},\ \Eprint
  {http://arxiv.org/abs/1201.6449} {arXiv:1201.6449 [hep-th]} \BibitemShut
  {NoStop}%
\bibitem [{\citenamefont {Chicherin}\ \emph {et~al.}(2016)\citenamefont
  {Chicherin}, \citenamefont {Drummond}, \citenamefont {Heslop},\ and\
  \citenamefont {Sokatchev}}]{Chicherin:2015edu}%
  \BibitemOpen
  \bibfield  {author} {\bibinfo {author} {\bibfnamefont {D.}~\bibnamefont
  {Chicherin}}, \bibinfo {author} {\bibfnamefont {J.}~\bibnamefont {Drummond}},
  \bibinfo {author} {\bibfnamefont {P.}~\bibnamefont {Heslop}}, \ and\ \bibinfo
  {author} {\bibfnamefont {E.}~\bibnamefont {Sokatchev}},\ }\href {\doibase
  10.1007/JHEP08(2016)053} {\bibfield  {journal} {\bibinfo  {journal} {JHEP}\
  }\textbf {\bibinfo {volume} {08}},\ \bibinfo {pages} {053} (\bibinfo {year}
  {2016})},\ \Eprint {http://arxiv.org/abs/1512.02926} {arXiv:1512.02926
  [hep-th]} \BibitemShut {NoStop}%
\end{thebibliography}%


\end{document}